\documentclass[12pt,preprintnumbers,superscriptaddress,nofootinbib]{revtex4}
\usepackage{multirow}
\usepackage{amssymb}
\usepackage{amsmath,graphicx}
\usepackage{dcolumn}
\usepackage{bm}
\usepackage{enumerate}
\usepackage{slashed}
\usepackage{epstopdf}
\usepackage[unicode]{hyperref}
\usepackage{braket}
\usepackage[usenames]{color}

\newcommand{\be}{\begin{equation}}
\newcommand{\ee}{\end{equation}}
\newcommand{\bea}{\begin{eqnarray}}
\newcommand{\eea}{\end{eqnarray}}
\newcommand{\ben}{\begin{enumerate}}
\newcommand{\een}{\end{enumerate}}
\newcommand{\bde}{\begin{widetext}}
\newcommand{\ede}{\end{widetext}}

\newcommand{\bc}{\begin{center}}
\newcommand{\ec}{\end{center}}

\begin{document}

\title{\boldmath Inflation with a class of concave inflaton potentials in Randall-Sundrum model}

\author{Ngo Phuc Duc Loc}
\email{locngo148@gmail.com}
\affiliation{Department of Theoretical Physics, University of Science, Ho Chi Minh City, 700000, Vietnam}
\affiliation{Vietnam National University, Ho Chi Minh City, 700000, Vietnam}

\begin{abstract}
We investigate inflation with a class of concave inflaton potentials of the form $\sim \phi^n$ $(0<n<1)$ in the Randall-Sundrum model with an infinite extra spatial dimension. We show that this class of models is much more in good agreement with observations compared to the standard inflation. We also find the range of the five-dimensional Planck scale ($M_5$) and show that large tensor-to-scalar ratios do not eliminate small-field inflation in braneworld cosmology.
	\end{abstract}
	\maketitle

\section{Introduction}

The mysterious large mass hierarchy puzzle, which is the huge discrepancy between the Planck scale and the electroweak scale, has attracted many theoretical physicists in the late 20th century. It addresses the profound question of why gravity is so weak. Braneworld models are motivated from such a puzzle and they offer a simple solution by a realization of the existence of extra spatial dimension(s) \cite{1,ADD,2}. We will work specifically with the Randall-Sundrum model which has an infinite extra dimension (RS2) \cite{3} in this paper because this model is phenomenologically more interesting in cosmology, although it cannot solve the hierarchy problem since there is only a single brane. 

Inflation is now widely believed to be a plausible mechanism in order to avoid fine-tuned initial conditions of the Universe, and it gives very good predictions, though with many parameters, that are compatible with observations. Primordial gravitational wave is the last important prediction which has not been confirmed yet. Future space-based gravitational wave detectors such as LISA may be able test this \cite{4}. In Section \ref{inflation}, we investigate inflation in RS2 model with a class of concave inflaton potentials that has been recently favored by Planck data \cite{5}. It is therefore interesting to see if this class is still valid in braneworld cosmology.

\section{Inflation in Randall-Sundrum model}\label{inflation}

Consider the following class of concave inflaton potentials
\begin{equation}\label{inflaton potential}
V(\phi)=\alpha^{4-n} \phi^{n},
\end{equation}
where $0<n<1$ and $\alpha>0$. There are two approximations that we will do, which are the usual slow-roll approximation and the high-energy approximation. While the slow-roll approximation is the must for inflation to happen, the high-energy approximation emphasizes the distinction between braneworld inflation and standard four-dimensional spacetime inflation. These two approximations must be maintained throughout inflation.

The modified Friedmann equation in Randall-Sundrum model is \cite{6}
\begin{equation}
H^2=\frac{8\pi}{3M_4^2}\rho\left(1+\frac{\rho}{2\lambda}\right),
\end{equation}
where $M_4$ is the usual four-dimensional Planck scale, $\rho$ is the energy density of the inflaton field, and $\lambda$ is the brane tension. Within the slow-roll approximation the kinetic term is sub-dominant and hence the energy density consists mainly of the inflaton potential, i.e. $\rho\simeq V$. In the low-energy limit, $V<<\lambda$, we recover the standard Friedmann equation. Therefore, the high-energy limit, $V>>\lambda$, characterizes braneworld cosmology. The brane tension is related to $M_4$ and $M_5$ (five-dimensional Planck scale) as \cite{6}
\begin{equation}
\lambda=\frac{3M_5^6}{4\pi M_4^2}.
\end{equation}
Because the initial inflaton potential must not exceed the Planck scale or otherwise quantum gravity effects will become important, we can deduce that $M_5<<M_4$ due to the high-energy approximation mentioned above. We will use this fact in subsequent discussions.

Under these two approximations, we have the following slow-roll parameters \cite{6}
\begin{equation}\label{slow-roll parameters}
\epsilon\simeq \frac{3M_5^6}{16\pi^2}\frac{(V')^2}{V^3}, \hspace{1cm}
\eta\simeq \frac{3M_5^6}{16\pi^2}\frac{V''}{V^2},
\end{equation}
where the primes denote derivatives with respect to $\phi$. The scalar spectral index, $n_s$, and the tensor-to-scalar ratio, $r$, are
\begin{equation}\label{observable quantities}
n_s\simeq 1-6\epsilon+2\eta,\hspace{1cm}
r\simeq 24\epsilon.
\end{equation}
The e-folding number is
\begin{equation}\label{e-folding number}
N\simeq \frac{-16\pi^2}{3M_5^6}\int_{\phi_i}^{\phi_f}\frac{V^2}{V'}d\phi,
\end{equation}
where $\phi_i$ and $\phi_f$ are the initial and final values of the inflaton field, respectively.

\begin{figure}
\begin{minipage}{\columnwidth}
\centering
\includegraphics[scale=1]{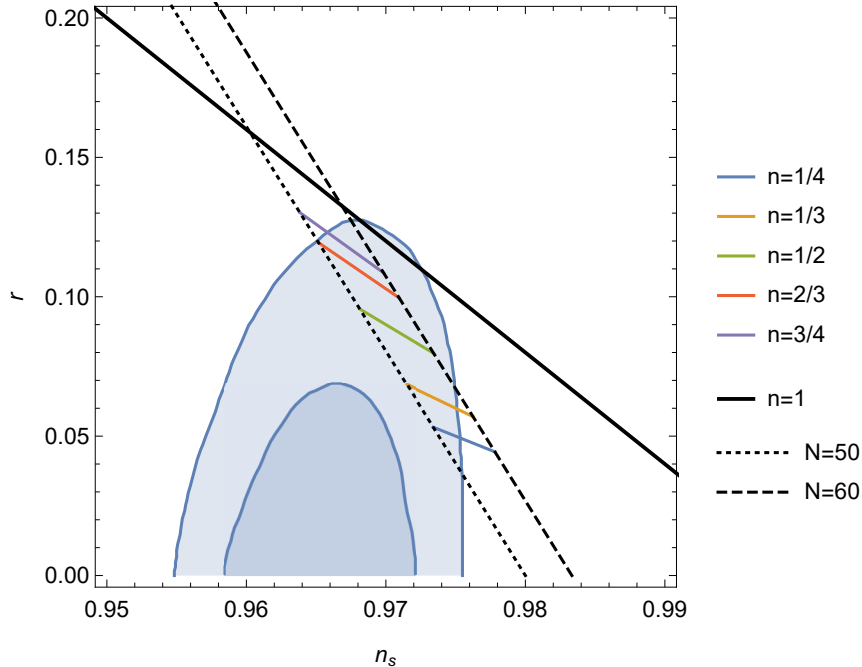}
\end{minipage}
\caption{Confrontation between the theoretical predictions of inflation in RS2 model with the Planck TT, TE, EE + low E + lensing data with $68\%$ and $95\%$ CL (shaded regions) \cite{5}. The convex inflaton potentials lie above the $n=1$ line, while the concave inflaton potentials lie below it.}
\label{RS2}
\end{figure}

From the condition of ending inflation, $\epsilon\approx 1$, we can find $\phi_f$. Substituting this value into the e-folding number we can find $\phi_i$ and hence $n_s, r$.
With the inflaton potential in Eq. \ref{inflaton potential}, we get the scalar spectral index as
\begin{equation}
n_s=1-\frac{2(2n+1)}{(n+2)N+n},
\end{equation}
and the tensor-to-scalar ratio is
\begin{equation}\label{tensor-to-scalar RS}
r=\frac{24n}{(n+2)N+n}.
\end{equation}
The Planck data suggests that $50\lesssim N \lesssim 60$ with 95\% confidence level \cite{5}, so we plot our results in Fig. \ref{RS2} for some typical values of $n$.

\begin{figure}
\begin{minipage}{\columnwidth}
\centering
\includegraphics[scale=1]{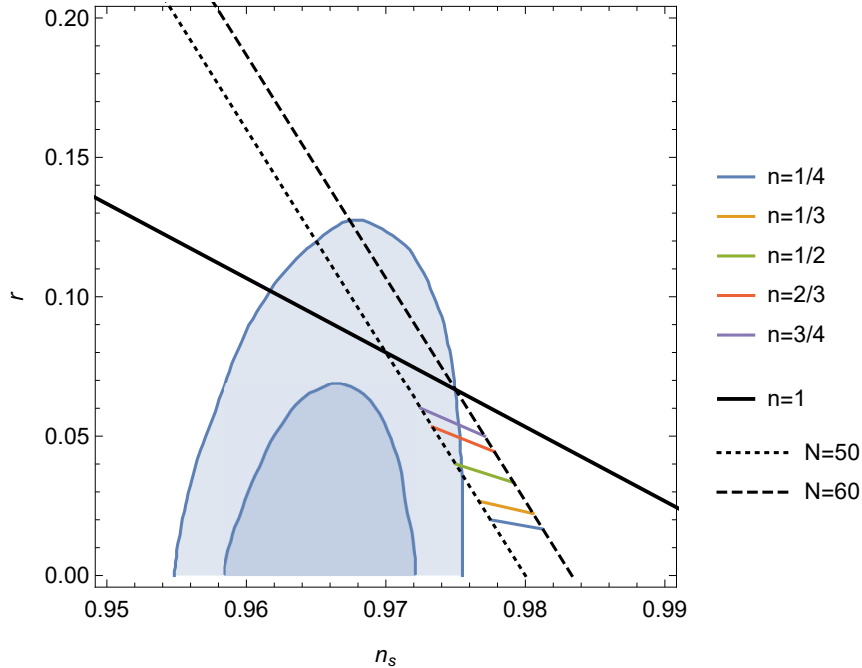}
\end{minipage}
\caption{Confrontation between the theoretical predictions of standard inflation with the Planck TT, TE, EE + low E + lensing data with $68\%$ and $95\%$ CL (shaded regions) \cite{5}. The convex inflaton potentials lie above the $n=1$ line, while the concave inflaton potentials lie below it.}
\label{standard}
\end{figure}

In standard inflation, we have the following results \cite{7}. The primordial tilt is
\begin{equation}
n_s=1-\frac{2+n}{2N},
\end{equation}
and the tensor-to-scalar ratio is
\begin{equation}
r= \frac{4n}{N}.
\end{equation}
We compare these results with the Planck data in Fig. \ref{standard} with some values of $n$.

From Fig. \ref{RS2} and Fig. \ref{standard}, we see that for each value of $n$ the scalar spectral index in RS2 model is smaller than the standard inflation case, but the tensor-to-scalar ratio is larger. Although the class of concave inflaton potentials of the form $\sim\phi^n$ in standard inflation is consistent with observations for some values of $n$ such as $n=2/3$ and $n=3/4$, these models are only valid if the number of e-folds $N$ is not large. On the other hand, the concave inflaton potentials of the form $\sim\phi^{1/2}$ and $\sim \phi^{2/3}$ are safely inside the allowed region for arbitrary numbers of e-folds in the range $50\lesssim N\lesssim 60$ in RS2 model. In Ref. \cite{7}, the authors showed that the inflaton potential of the form $\sim\phi^{2/3}$ is physically motivated from the context of string inflation, while in our work this class of models is just a phenomenological approach to achieve desired results that are in good agreement with observations. We admit that the arbitrariness of inflaton potentials is a weakness of the inflationary theory, and that finding a plausible physical mechanism to actually generate the proper potential is also important and needs to be studied further.

In addition, we also want to comment on the relationship between the inflaton field excursion and the tensor-to-scalar ratio in RS2 model. This can easily be obtained from Eq. \ref{e-folding number}, Eq. \ref{observable quantities}, and Eq. \ref{slow-roll parameters} as
\begin{equation}
\Bigg|\frac{d\phi}{dN}\Bigg|=\frac{M_5^3}{ 8\sqrt{2}\pi}\sqrt{\frac{r}{V}}.
\end{equation}
This is the Lyth bound in RS2 model. In standard inflation, the Lyth bound does not depend explicitly on the inflaton potential, and if the inflaton field excursion is super-Planckian (large-field inflation), then the tensor-to-scalar ratio is large and primordial gravitational wave is an interesting observational signature of large-field inflationary models \cite{Starobinsky, 8}. In RS2 model, however, this is not the case. The field excursion now depends on the inflaton potential itself, so that the detailed analysis is therefore model-dependent. Nevertheless, we can still easily see that the field excursion can be sub-Planckian (small-field inflation) from the qualitative viewpoint. The reason is that the inflaton potential must not exceed the 4D Planck scale (i.e. $V\lesssim M_4^4$)  and $M_5<<M_4$ due to the high-energy approximation so that the field excursion is smaller than the 4D Planck scale $\Delta\phi<M_4\sqrt{r}$. Therefore, large tensor-to-scalar ratios do not necessarily imply large-field inflation, and observations of gravitational wave do not eliminate small-field inflation in braneworld cosmology. This remark can also be seen more explicitly by looking at the initial value of the inflaton field of $\phi^n$ model
\begin{equation}
\phi_i^{n+2}=\frac{3M_5^6}{16\pi^2}\frac{n[(n+2)N+n]}{\alpha^{4-n}}.
\end{equation}
We can freely choose the appropriate parameters of $M_5$ and $\alpha$, provided that the initial inflaton potential is smaller than $M_4^4$, to demand that the initial inflaton field is sub-Planckian and hence small-field inflation is established. Meanwhile, the tensor-to-scalar ratio (Eq. \ref{tensor-to-scalar RS}), which is independent of these two parameters, is large in this model.

On the other hand, the lower bound of $M_5$ is determined from the accuracy of the Newtonian gravitational potential at small distance, because the Randall-Sundrum model with an infinite extra dimension predicts the modification of such a potential as \cite{3,6}
\begin{equation}
V(r)=-\frac{Gm_1m_2}{r}\left(1+\frac{1024\pi^2M_4^4}{r^2M_5^6}\right).
\end{equation}
The Newtonian gravitational potential was tested to an incredibly small distance of about $52\mu m$ \cite{9}. The lower bound of the 5D Planck scale is therefore
\begin{equation}\label{lower}
M_5^6>>\frac{1024 \pi^2 M_4^4}{r^2}\approx 3,23.10^{57} GeV^6.
\end{equation}
So overall we have the range
\begin{equation}
3,84.10^9 GeV<<M_5<<1,22.10^{19} GeV.
\end{equation}
This range could be useful for future investigations.

\section{Conclusion}

In this paper, we showed that the concave inflaton potentials of the form $\sim \phi^n$ in the Randall-Sundrum type II model are consistent with observations if $1/3\lesssim n\lesssim 3/4$. Inflation in RS2 model generally predicts smaller scalar spectral indices and larger tensor-to-scalar ratios compared to standard inflation. However, large tensor-to-scalar ratios do not strictly imply large-field inflation in this model since the inflaton field rolls more slowly. Our possible future work will be investigating how the braneworld cosmology might affect certain processes in the early Universe such as electroweak baryogenesis \cite{10}.

\begin{acknowledgements}
I thank Dr. Vo Quoc Phong for his support and encouragement.
\end{acknowledgements}


\begin{thebibliography}{9}

\bibitem{1} Nima Arkani-Hamed, Savas Dimopoulos, Gia Dvali, Phys.Lett.B \textbf{429}: 263-272 (1998).

\bibitem{ADD} I. Antoniadis, N. Arkani-Hamed, S. Dimopoulos, G. Dvali, Phys.Lett.B \textbf{436}: 257-263 (1998).

\bibitem{2} L. Randall and R. Sundrum, Phys. Rev. Lett. \textbf{83}, 3370 (1999).

\bibitem{3} L. Randall and R. Sundrum, Phys. Rev. Lett. \textbf{83}, 4690 (1999).


\bibitem{4} P. A. Seoane et al. (eLISA Collaboration), The gravitational universe, arXiv:1305.5720

\bibitem{5} Y. Akrami et al. (Planck Collaboration), Planck 2018 results. X. Constraints on inflation, arXiv:1807.06211v2.


\bibitem{6} Roy Maartens, David Wands, Bruce A. Bassett, and Imogen P. C. Heard, Phys. Rev. D \textbf{62}, 041301(R) (2000).



\bibitem{7} Eva Silverstein and Alexander Westphal, Phys. Rev. D \textbf{78}, 106003 (2008).


\bibitem{Starobinsky} A. A. Starobinsky, Sov. Astron. Lett. \textbf{11}, 133 (1985).



\bibitem{8} David H. Lyth, Phys. Rev. Lett. \textbf{78}, 1861 (1997).

\bibitem{9} J. G. Lee, E. G. Adelberger, T. S. Cook, S. M. Fleischer, and B. R. Heckel, Phys. Rev. Lett. \textbf{124}, 101101 (2020).


\bibitem{10} Vo Quoc Phong, Phan Hong Khiem, Ngo Phuc Duc Loc, and Hoang Ngoc Long, Phys. Rev. D \textbf{101}, 116010 (2020).

\end{thebibliography}
\end{document}